\documentclass[prd,aps,tightenlines,preprint,nofootinbib,superscriptaddress]{revtex4}
\usepackage{epsfig}
\usepackage{amsmath}
\input{epsf}

\begin{document}


\vspace*{2cm}
\title{ Transverse momentum dependent quark distributions and polarized Drell-Yan processes }

\author{Jian Zhou}
\affiliation{School of Physics, Shandong University, Jinan, Shandong 250100, China}
\affiliation{Nuclear Science Division, Lawrence Berkeley National Laboratory, Berkeley, CA 94720}
\author{Feng Yuan}
\affiliation{Nuclear Science Division, Lawrence Berkeley National Laboratory, Berkeley, CA 94720}
\affiliation{RIKEN BNL Research Center, Building 510A, Brookhaven National Laboratory, Upton, NY 11973}
\author{Zuo-Tang Liang}
\affiliation{School of Physics, Shandong University, Jinan, Shandong 250100, China}

\begin{abstract}
We study the spin-dependent
quark distributions at  large transverse momentum. We derive their transverse momentum behaviors
in the collinear factorization approach in this region.
We further calculate the angular distribution of the Drell-Yan lepton pair production with polarized beams
and present the results in terms of the collinear twist-three quark-gluon correlation functions.
In the intermediate transverse momentum region,
we find that the two approaches: the collinear factorization
and the transverse momentum dependent factorization approaches are consistent
in the description of the lepton pair angular distributions.
\end{abstract}

\maketitle

\section{Introduction}
Spin dependent semi-inclusive hadronic processes have attracted much interest
from both  experiment and theory sides in recent years.
These processes provide us more opportunities to study the Quantum Chromodynamics (QCD)
and internal structure of the hadrons,
as compared to the inclusive hadronic processes or spin averaged processes.
Measurements have been made in different reactions.
In particular, the single transverse spin asymmetry (SSA) phenomena
observed in various hadronic processes\cite{E704-Bunce,hermes,dis,phenix,star,brahms,belle}
have stimulated remarkable theoretical developments
\cite{Liang:1989mb,review123,Ji:1993vw,Sivers:1990fh,Collins:1992kk,Anselmino:1994tv,Arnold:2008kf,
Bacchetta:2006tn,Boer:1999mm,Brodsky:2002cx,Collins:2002kn,Ji:2002aa,{Cherednikov:2007tw},
Boer:2003cm,JiMaYu04,ColMet04,Mulders:1995dh,Boer:1997nt,Liang:2006wp}.
Among the theoretical developments, two
approaches in the QCD framework
have been most explored: the higher twist collinear factorization approach
\cite{qiu,Efremov,Eguchi:2006qz,new}
and the transverse momentum dependent (TMD) approach
\cite{Sivers:1990fh,Collins:1992kk,Anselmino:1994tv,Arnold:2008kf,
Bacchetta:2006tn,Boer:1999mm,Brodsky:2002cx,Collins:2002kn,Ji:2002aa,{Cherednikov:2007tw},
Boer:2003cm,JiMaYu04,ColMet04,Mulders:1995dh,Boer:1997nt,Liang:2006wp}.  In these two approaches,
the spin-dependent differential cross sections can be calculated in terms of the
collinear twist three quark-gluon correlation functions in the
collinear factorization formalism  and the TMD distributions in the TMD factorization
approach. Such functions generalize the
original Feynman parton picture, where the partons only carry longitudinal
momentum fraction of the parent hadron. They will certainly
provide more information on hadron structure.
Further study has shown that the transverse momentum dependence of the
naive-time-reversal-odd TMD distributions which are responsible for the SSAs
can be calculated and expressed in terms of the collinear twist-three correlation functions
at the large transverse momentum. By using these results, it was shown
that the above two approaches are consistent in the intermediate transverse
momentum region where both apply~\cite{jqvy,{Zhou:2008fb},Yuan:2009dw}.
In this paper, we will extend these studies to more general Drell-Yan
processes, in particular the lepton angular distributions in polarized nucleon-nucleon
scattering.

The single transverse spin asymmetry in the Drell-Yan lepton
pair production process has been used as an example to demonstrate
the consistency between these two approaches~\cite{jqvy}, where
the single transverse spin asymmetry is represented as
a correlation between the lepton pair transverse momentum $q_\perp$
and the transverse polarization vector $S_\perp$. For this contribution, the transverse
spin dependent differential cross section is proportional to
$d\sigma(S_\perp)\propto \epsilon^{ij}S_\perp^iq_\perp^j$. The transverse
momentum of the lepton pair is also the transverse momentum
of the virtual photon which decays into the lepton pair in the Drell-Yan
process. Therefore, this SSA is related to the quark Sivers function,
and is the only contribution from the quark-antiquark channel.
However, if we further study the lepton angular distribution in the Drell-Yan
process~\cite{Drell:1970wh}, it will open more contributions to the single spin asymmetries,
as well as other spin dependent observables~\cite{Collins:1977iv,{Lam:1978pu}}.
More recently, by analyzing the general Lorentz structure of the hadronic tensor,
the complete spin and transverse momentum dependent angular distribution of lepton pair
has been presented in Ref.~\cite{Arnold:2008kf}, and 48 structure functions will contribute.
One can calculate some of them which are leading power in the context of the TMD factorization
and relate them to the TMD distributions~\cite{Arnold:2008kf}.

In this paper, we will study the angular distribution of the lepton pair
in the polarized Drell-Yan process at large transverse momentum of the lepton pair.
The relevant calculations are carried out in the collinear factorization framework.
We mainly focus on the single spin asymmetry $A_{UT}$ and double spin asymmetry $A_{LT}$,
taking into account the contributions from the twist-three quark-gluon correlation
functions from one of the incident hadrons. We will limit ourselves up to this order in the
calculations.
For the single spin asymmetry $A_{UT}$, the corresponding twist-three quark-gluon correlation
functions are those studied in~\cite{qiu,new,jqvy,Zhou:2008fb,{Eguchi:2006qz}} and the
calculations will be similar. On the other
hand, for the $A_{LT}$ asymmetry, more general quark-gluon correlation functions
will contribute and the calculations will be different from those in~\cite{qiu,new,jqvy,Zhou:2008fb,{Eguchi:2006qz}}.
Following the same procedure as that in Refs.~\cite{jqvy,Yuan:2009dw,{Zhou:2008fb}}, we
will compare the predictions from the two formalisms
and check their consistency.

To pursue this aim, we will calculate the TMD quark distributions at large
transverse momentum $k_\perp \gg \Lambda_{\rm QCD}$, and express them
in terms of the collinear correlation functions.
There are eight leading order TMD quark distributions
\cite{Mulders:1995dh}: three $k_\perp$-even TMD distributions
$f_1(x,k_\perp), g_{1L}(x,k_\perp), h_1(x,k_\perp)$;
two naive-time-reversal-odd TMD quark distributions
$f_{1T}^\perp(x,k_\perp)$ (Sivers function), $h_1^\perp(x,k_\perp)$ (Boer-Mulders function);
and three naive-time-reversal-even but $k_\perp$-odd TMD quark distributions
$g_{1T}(x,k_\perp)$,  $h_{1L}(x,k_\perp)$,  $h^\perp_{1T}(x,k_\perp)$.
The transverse momentum dependence of these TMD quark distributions
can be calculated from perturbative QCD in the collinear factorization
framework.
For the $k_\perp$-even TMD quark distributions, the results are well-known, and
can be expressed in terms of the integrated leading-twist parton
distributions (see, e.g., ~\cite{JiMaYu04}).
The naive-time-reversal-odd TMD quark distributions (the Sivers function
and Boer-Mulders function) have also been calculated~\cite{jqvy,{Zhou:2008fb}}.
In this paper, we will extend these calculations to the two naive-time-reversal-even
but $k_\perp$-odd TMD quark distributions $g_{1T}$ and $h_{1L}$. These
results will depend on the the novel twist three distributions $\tilde{g}(x)$,
$\tilde{h}(x)$~\cite{Boer:2003cm,Eguchi:2006qz}, and the general twist-three
quark-gluon correlation functions $G_D$, $\tilde G_D$, $H_D$, and $E_D$~\cite{Jaffe:1991ra}.
The last TMD quark distribution $h^\perp_{1T}(x,k_\perp)$ will involve twist-four
quark-gluon correlation functions. We will not discuss it in this paper.

The rest of paper is organized as follows. In Sec.~II, we
give a brief review on the twist-three quark-gluon collinear
correlation matrix elements and discuss the relation between them
and the TMD distributions. General feature of the TMD quark
distributions at large transverse momentum will be presented in
Sec.~III. In Sec.~IV, we derive the
naive-time-reversal-even TMD quark distributions
$g_{1T}(x,k_\perp)$, $h_{1L}(x,k_\perp)$ in the twist-three
quark-gluon correlation approach. In Sec.~V, we calculate the
relevant polarized Drell-Yan differential
cross section using the same collinear factorization and compare
to the results from the TMD factorization. We conclude the paper
in Sec.~VI.

\section{Twist-3 correlation matrix elements and TMD distributions}

In order to extract more information on hadron structure, various spin dependent and/or
transverse momentum dependent
parton correlation functions have been introduced based on the QCD factorization theorem.
They are universal between SIDIS and Drell-Yan processes (up to a sign for the naive-time-reversal-odd TMD parton
distributions) and can be pinned down by a complete set of experiments.
It has been shown that there exits interesting connections between the twist-three collinear
functions and the naive-time-reversal-odd TMD quark distributions~\cite{Boer:2003cm,Eguchi:2006qz}.
In this section, we will review the general property of the collinear correlation functions and introduce
two novel twist-three functions $\tilde{g}, \tilde{h}$ \cite{Boer:2003cm}.
We will further explore their relations to the naive-time-reversal-even
but $k_\perp$-odd TMD quark distributions $g_{1T}$ and $h_{1L}$.

Let us start by introducing the following collinear quark-antiquark correlation matrix:
\begin{eqnarray}
{ \hat{M}}_{\alpha \beta } (x )\equiv \int \frac{dy^-}{2 \pi} e^{-ixP^+ y^-}
\langle P,S | \bar{\psi}_\beta(y^-) \psi_\alpha(0) | P,S \rangle  ,
\end{eqnarray}
where $P,S$ are the hadron momentum and spin, respectively, and we have suppressed the
light-cone gauge links between different fields.  The hadron momentum $P^\mu$ is proportional to the light cone
vector $p^\mu=(1^+,0^-,0_\perp)$, whose conjugate light-cone vector is $n=(0^+,1^-,0_\perp)$.
$x$ is the momentum fraction of the hadron carried by the quark.
Up to twist-three level, the above matrix can be expanded as
~\cite{Jaffe:1991ra},
\begin{eqnarray}
{\hat { M}}(x )&=&\frac{1}{2} \left [ f_1(x) p\!\!\!/ + g_1(x)\lambda \gamma_5 p\!\!\!/ +h_1(x)\gamma_5 S\!\!\!/_\perp p\!\!\!/
\right ]
\nonumber \\
&+&\frac{M}{2P^+} \left [ e(x) {\rm 1} +g_T(x) \gamma_5 S\!\!\!/_\perp + h_L(x)\frac{\lambda}{2} \gamma_5 [p\!\!\!/,n\!\!\!/]
\right ] \ ,
\end{eqnarray}
where $\lambda$ represents the helicity for the nucleon for the longitudinal polarized
nucleon, $S_\perp$ is the transverse polarization vector, and $M$ the hadron mass.
The first three are the leading-twist quark distributions: spin average $f_1$, longitudinal spin $g_1$,
and quark transversity $h_1$ distributions. The twist-three quark distributions: $e(x)$, $g_T(x)$, and $h_L(x)$
do not have simple interpretations, and belong to more general quark-gluon
correlation functions~\cite{Jaffe:1991ra}.
These correlation functions can be defined through the following matrix~\cite{Ellis:1982cd,Jaffe:1991ra,Qiu:1990xxa},
\begin{eqnarray}
{\hat{ M} }_{D \alpha \beta }^\mu(x,x_1)\equiv \int \frac{dy^-}{2 \pi} \frac{dy_1^-}{2\pi} e^{-ixP^+ y^-} e^{i(x_1-x)P^+y^-_1}
\langle P,S | \bar{\psi}_\beta(y^-)  i D_\perp^\mu(y_1^-)\psi_\alpha(0) | P,S \rangle  \ ,
\end{eqnarray}
where we have adopted the covariant derivative as $i D_\perp^\rho=
i \partial^\rho + g A_\perp^\rho$. The expansion of the above
matrix contains the following four twist-three quark-gluon
correlation functions,
\begin{eqnarray}
{\hat{M}}_{D }^\mu(x,x_1)&=& \frac{M}{2P^+} \left [ G_D(x,x_1)i\epsilon^{\mu\nu}_\perp S_{\perp \nu} p\!\!\!/
+\tilde{G}_D(x,x_1)S_\perp^\mu \gamma_5 p\!\!\!/ \right ]
\nonumber \\
&+& \frac{M}{2P^+} \left [
H_D(x,x_1) \lambda \gamma_5 \gamma_\perp^\mu p\!\!\!/ +E_D(x,x_1) \gamma_\perp^\mu p\!\!\!/ \right ] \ .
\end{eqnarray}
By imposing the hermiticity, parity and time-reversal invariance, we will have the following constrains,
\begin{eqnarray}
&&G_D(x,x_1)=-G_D(x_1,x) , \  \tilde{G}_D(x,x_1)=\tilde{G}_D(x_1,x),
\nonumber \\
&&H_D(x,x_1)=H_D(x_1,x) , \ E_D(x,x_1)=-E_D(x_1,x)  \ ,
\end{eqnarray}
and these functions are real.
As mentioned, the twist-three quark distribution $g_T(x)$, $e(x)$, and $h_L(x)$ can
be expressed in terms of the above quark-gluon correlation functions~\cite{Jaffe:1991ra,Ratcliffe:1985mp},
\begin{eqnarray}
&&g_T(x)=\frac{1}{x} \int dx_1 \left [ G_D(x,x_1) + \tilde{G}_D(x,x_1) \right ] \ ,
\\&&
h_L(x)= \frac{2}{x} \int dx_1 H_D(x,x_1) \ ,
\\&&
e(x)= \frac{2}{x} \int dx_1 E_D(x,x_1) \ .
\end{eqnarray}
Therefore, $G_D$, $\tilde G_D$, $H_D$, and $E_D$ functions are more fundamental, which
becomes evident when we study the scale evolution for the twist-three quark
distributions, and the next-to-leading order perturbative corrections to the relevant
cross sections~\cite{Ratcliffe:1985mp,{Ji:2001bm}}.

In terms of the twist expansion, of course, $D$-type correlations are not the only
ones at the twist-three level. One can also define a set of the
$F$-type twist-3 correlation matrix elements,
\begin{eqnarray}
\hat{M}_{F \alpha \beta  }^\mu(x,x_1)\equiv \int \frac{dy^-}{2 \pi} \frac{dy_1^-}{2\pi} e^{-ixP^+ y^-} e^{i(x_1-x)P^+y^-_1}
\langle P,S | \bar{\psi}_\beta(y^-)  g F_{+\perp}^\mu(y_1^-)\psi_\alpha(0) | P,S \rangle  \ .
\end{eqnarray}
Again, the expansion of the above matrix defines the following $F$-type quark-gluon
correlation functions,
\begin{eqnarray}
\hat{M}_{F }^\mu(x,x_1)&=& \frac{M}{2} \left [ T_F(x,x_1)\epsilon^{\nu\mu}_\perp S_{\perp \nu} p\!\!\!/
+\tilde{T}_F(x,x_1)iS_\perp^\mu \gamma_5 p\!\!\!/ \right ]
\nonumber \\
&+& \frac{M}{2} \left [
\tilde{T}_F^{(\sigma)}(x,x_1) i\lambda \gamma_5 \gamma_\perp^\mu p\!\!\!/ +T_F^{(\sigma)}(x,x_1) i\gamma_\perp^\mu p\!\!\!/ \right ] \ ,
\end{eqnarray}
where, for convenience, we have used different normalization factors for
$T_F(x,x_1)$ and $T_F^{(\sigma)}(x,x_1)$ as compared
to Ref.~\cite{jqvy,{Zhou:2008fb}}, with a relative factor $2\pi M$.
Similarly, the parity and time-reversal invariance implies,
\begin{eqnarray}
&&T_F(x,x_1)=T_F(x_1,x) , \  \tilde{T}_F(x,x_1)=-\tilde{T}_F(x_1,x),
\nonumber \\
&&\tilde{T}_F^{(\sigma)}(x,x_1)=-\tilde{T}_F^{(\sigma)}(x_1,x) ,\  T_F^{(\sigma)}(x,x_1)=T_F^{(\sigma)}(x_1,x) \ .
\end{eqnarray}
The $F$-type correlation functions are usually regarded as an alternative but
not independent functions in the calculations to the inclusive DIS
structure functions, such as $g_T$ structure function~\cite{Ji:2001bm}.
On the other hand, it has been found that the $F$-type correlation functions
are more relevant for the single transverse spin asymmetry, and have been
intensively studied~\cite{qiu,new,{Eguchi:2006qz}}.
By using the equation of motion, these two types of the correlation
functions can be related to each other~\cite{{Boer:2003cm},Eguchi:2006qz},
\begin{eqnarray}
&&G_D(x,x_1)=P \frac{1}{x-x_1} T_F(x,x_1),
\\&&
\tilde{G}_D(x,x_1)=P \frac{1}{x-x_1} \tilde{T}_F(x,x_1)+\delta(x-x_1)\tilde{g}(x),
\\&&
E_D(x,x_1)=P \frac{1}{x-x_1} T_F^{(\sigma)}(x,x_1),
\\ &&
H_D(x,x_1)=P \frac{1}{x-x_1} \tilde{T}_F^{(\sigma)}(x,x_1)+\delta(x-x_1)\tilde{h}(x).
\end{eqnarray}
where $P$ stands for the principal value, and $\tilde{g}, \tilde{h} $ are given by~\cite{Boer:2003cm,Eguchi:2006qz},
\begin{eqnarray}
&&\int \frac{dy^-}{2 \pi} e^{-ixP^+ y^-}
\langle P,S | \bar{\psi}(y^-)  \left( iD_\perp^\mu-ig\int_0^\infty d\zeta^-F^{+\perp}(\zeta^-)\right)
\psi(0) | P,S \rangle \nonumber\\
&&~~~=\frac{M}{2} \left [ \tilde{g}(x) S_\perp^\mu \gamma_5 p\!\!\!/ + \tilde{h}(x) \lambda \gamma_5 \gamma_\perp^\mu p\!\!\!/ \right ] \ .
\end{eqnarray}
From the above results, we find that indeed the $F$-type and $D$-type correlation
functions are not completely independent, and they form an over-complete
set of functions. However, we still need $\tilde g$ and $\tilde h$ to completely describe
the associated physics at this order, in particular, for the calculation performed in this paper.
In the real calculations, we can either use $D$-type or $F$-type
plus $\tilde g$ and $\tilde h$ as a complete set of twist-three
functions.

In the following, we will further reveal the physical meaning of
$\tilde g$ and $\tilde h$, and build the connection between
them and the transverse momentum dependent quark distributions.
The TMD parton distributions are important generalization of
the conventional Feynman parton distributions. Because
of additional dependence on the transverse momentum of
partons, these distributions open more opportunities to study
the partonic structure in nucleon. The nontrivial correlations
between the parton transverse momentum and the polarization
vectors of the parent nucleon or the quark itself provide
novel consequence in the transverse component in the hadronic
processes, for example, the single transverse spin
asymmetry. Of course, upon integral over transverse momentum,
these TMD parton distributions will naturally connect to the
leading-twist and higher-twist parton distributions.
 In this paper, we will focus on the TMD quark distributions,
which are relevant to the Drell-Yan lepton pair production.
The TMD quark distributions can be
defined through the following matrix~\cite{Collins:1981uw,Collins:2002kn,Ji:2002aa},
\begin{equation}
{\hat{\cal M}}_{\alpha \beta }(x,k_\perp)=
\int\frac{dy^-d^2y_\perp}{(2\pi)^3} e^{-ixP^+\cdot
y^-+i\vec{k}_\perp\cdot \vec{y}_\perp}\langle
PS|\overline{\psi}_\beta(y^-,y_\perp) {\cal L}_v^\dagger(y^-,y_\perp) {\cal L}_v(0)
\psi_\alpha(0)|PS\rangle \ ,
\end{equation}
where $x$ is the longitudinal momentum fraction and $k_\perp$ the transverse
momentum carried by the quark. The gauge link ${\cal L}_v$ is along the
direction represented by $v$ which is conjugated to $p$. In the
case that we need to regulate the light-cone singularities, we will
use an off-light-cone vector: $v^-\gg v^+$ and $v_\perp=0$, and further
define $\zeta^2=(v\cdot P)^2/v^2$.
Compared with the integrated parton distributions definition in the above,
we find that the two quark fields are not only
separated by light-cone distance $\xi^-$, but also by the
transverse distance $\xi_\perp$, which is conjugate to the
transverse momentum of the quark $k_\perp$~\cite{Collins:1981uw}. Because of this
difference, the additional transverse gauge link for the TMD parton distributions has
to be contained in order to make the above quark matrix gauge invariant~\cite{Ji:2002aa},
and the gauge link direction depends on the process~\cite{Collins:2002kn,{Ji:2002aa}}.
Since we will study the Drell-Yan lepton pair production in this paper, in the following
we will adopt the TMD definition for this process and the gauge link will go to
$-\infty$~\cite{Collins:2002kn,{Ji:2002aa}}.
The gauge link plays an essential role in the naive time-reversal-odd TMD parton distributions.

The leading order expansion of the quark distribution matrix
${\cal M}$ contains eight quark distributions
\cite{Mulders:1995dh,Boer:1997nt,{Bacchetta:2006tn}},
\begin{eqnarray}
{\hat{\cal M}}&=&\frac{1}{2}\left[f_1(x,k_\perp)\not\!
p+\frac{1}{M}h_1^\perp(x,k_\perp)\sigma^{\mu\nu}k_\mu p_\nu+g_{1L}(x,k_\perp) \lambda \gamma_5\not\! p  \right.\nonumber\\
&& +\frac{1}{M}g_{1T}(x,k_\perp)\gamma_5\not\!
p(\vec{k_\perp}\cdot \vec{S}_\perp)+\frac{1}{M}h_{1L}\lambda
i\sigma_{\mu\nu}\gamma_5 p^\mu k_\perp^\nu +h_1(x,
k_\perp)i\sigma_{\mu\nu}\gamma_5 p^\mu S_\perp^\nu
\nonumber\\
&&\left.+\frac{1}{M^2}h_{1T}^\perp(x,k_\perp)i\sigma_{\mu\nu}\gamma_5
p^\mu \left(\vec{k}_\perp\cdot\vec{S}_\perp
k_\perp^\nu-\frac{1}{2}\vec{k}_\perp^2S_\perp^\nu\right)
+\frac{1}{M}f_{1T}^\perp(x,k_\perp)\epsilon^{\mu\nu\alpha\beta}\gamma_\mu
p_\nu k_\alpha S_\beta \right]\ . \label{tmdpar}
\end{eqnarray}
Out of the eight TMD distributions, three of them are associated with the
$k_\perp$-even structure: $f_1(x,k_\perp)$, $g_{1L}(x,k_\perp)$, and
$h_1(x,k_\perp)$. They are simple extension of the above
integrated quark distributions. The other five distributions are
associated with the $k_\perp$-odd structures, and hence vanish
when $k_\perp$ are integrated out for ${\cal M}_{\alpha\beta}$.
For an unpolarized nucleon target, one
can introduce the unpolarized quark distribution $f_1(x,
k_\perp)$ and naive-time-reversal-odd transversely-polarized quark
distribution $h_1^\perp(x,k_\perp)$, the Boer-Mulders function.
For a longitudinally-polarized
nucleon, one introduces a longitudinally-polarized quark
distribution $g_{1L}(x, k_\perp)$ and a transversely-polarized
distribution $h_{1L}^\perp(x, k_\perp)$. Finally, for a
transversely-polarized nucleon, one introduces a quark spin-independent
distribution $f^\perp_{1T}(x, k_\perp)$, the Sivers function,
and a longitudinally-polarized
quark polarization $g_{1T}(x, k_\perp)$, a symmetrical
transversely-polarized quark distribution $h_{1}(x, k_\perp)$ and
an asymmetric transversely-polarized quark distribution
$h_{1T}^\perp(x, k_\perp)$.

If we weight the integral of ${\cal M}_{\alpha\beta}$
with linear dependent transverse momentum, the $k_\perp$-odd
quark distributions will lead to the higher-twist quark-gluon
correlation functions~\cite{Boer:2003cm}. Four of them will correspond
to the four quark-gluon correlation functions introduced above, including
$f_{1T}^\perp$, $h_1^\perp$,  $g_{1T}$,  $h_{1L}$. The last one $h_{1T}^\perp$, as we mentioned,
will correspond to a twist-four correlation function.
First, the two naive-time-reversal-odd quark distributions $f_{1T}^\perp$
and $h_1^\perp$ lead to the following quark-gluon correlations\footnote{These relations
Eqs.~(19-22) are valid at the leading order in perturbative expansion which we will use
in this paper. They may differ from these forms at higher orders.}~\cite{Boer:2003cm,{Ma:2003ut}},
\begin{eqnarray}
&&T_F(x,x)=\int \frac{d^2 \vec{k}_\perp}{2\pi} \frac{\vec{k}_\perp^2}{M^2}f_{1T}^\perp|_{\rm DY}(x,k_\perp) \ ,
\\&&
T_F^{(\sigma)}(x,x)=\int \frac{d^2 \vec{k}_\perp}{2\pi} \frac{\vec{k}_\perp^2}{M^2}h_{1}^\perp|_{\rm DY} (x,k_\perp)\ ,
\end{eqnarray}
where the TMD quark distributions follow their definitions in the Drell-Yan process.
Similarly, the two naive-time-reversal-even but $k_\perp$-odd quark
 distributions $g_{1T}$ and $h_{1L}$ can be related to the following twist-three matrix element,
\begin{eqnarray}
&&\tilde{g}(x)=\int d^2 \vec{k}_\perp \frac{\vec{k}_\perp^2}{2M^2}g_{1T}(x,k_\perp) \ ,
\\&&
\tilde{h}(x)=\int d^2 \vec{k}_\perp \frac{\vec{k}_\perp^2}{2M^2}h_{1L} (x,k_\perp)\ .
\end{eqnarray}
Here, because they are naive-time-reversal-even distributions, they will not
change sign between DIS and Drell-Yan processes.
In summary, the four $k_\perp$-odd TMD quark distributions correspond to the
four twist-three quark-gluon correlation functions introduced above Eqs.~(12)-(15).
From these relations, we can further study the scale evolutions for the above twist
three correlations~\cite{{Vogelsang:2009pj},kang,Zhou:2008mz}.

The above relations are only one aspect of the connections between the TMD quark
distributions and higher-twist quark-gluon correlation functions. In the following section,
we will explore another aspect, i.e., the large transverse momentum behavior for
the TMD quark distributions in terms of the collinear leading-twist or
twist-three quark-gluon correlation functions.

\section{quark distributions at large transverse momentum }

When the $k_\perp$ is of the order of $\Lambda_{QCD}$, the TMD parton distribution functions are entirely
non-perturbative objects. However, the transverse momentum dependence
can be calculated in the perturbative QCD and related to
the collinear matrix elements as long as the $k_\perp$ is much larger than $\Lambda_{QCD}$, the
typical nonperturbative scale.
The collinear matrix elements are the relevant integrated
leading-twist parton distributions or higher-twist quark-gluon
correlation functions. For example, for the three $k_\perp$-even
quark distributions, they will depend on the integrated leading-twist
parton distributions. However, for the $k_\perp$-odd quark distributions,
they depend on the twist-three (or twist-four) quark-gluon correlation
functions.
In general, we will have the following expression for the quark distributions
at large transverse momentum~\cite{JiMaYu04}\footnote{This is not
a rigorous factorization formula. However, we shall be able to construct a QCD factorization
formalism in the impact parameter $b$-space for the TMD distributions
~\cite{{Collins:1981uw},JiMaYu04,Collins:1984kg}.},
\begin{equation}
q(x,k_\perp)|_{k_\perp\gg \Lambda_{\rm QCD}}=\frac{1}{(k_\perp^2)^n}\int \frac{dx'}{x'}
f_{i}(x')\times {\cal H}_{q/i}(x;x') \ ,
\end{equation}
where $q(x,k_\perp)$ represents the TMD quark distribution we are interested,
$f_i$ represents the integrated quark distribution for the $k_\perp$-even TMDs,
and higher twist quark-gluon correlation function for the $k_\perp$-odd TMDs.
For the latter case, $x'$ should be understood as two variables for the
twist-three quark-gluon correlation functions as we discussed in the last section.
The overall power behavior $1/(k_\perp^2)^n$ can be analyzed by the
power counting rule~\cite{Brodsky:1973kr}. The hard coefficient ${\cal H}_{q/i}(x;x')$ is calculated
from perturbative QCD. In this paper, we will show the one-gluon radiation
contribution to this hard coefficient.

The $k_\perp$-even TMD quark distribution functions,
$f_1(x,k_\perp)$,  $g_{1L}(x,k_\perp)$, and $h_1(x,k_\perp)$ can be
calculated from the associated integrated quark distributions~\cite{JiMaYu04}\footnote{Mixing with the
gluonic contributions will have to be taken into account for $f_1$ and $g_1$ distributions. In this paper,
we will not discuss these contributions.}.
For the non-singlet contributions, they are expressed as~\cite{JiMaYu04},
\begin{eqnarray}
f_1(x_B,k_\perp)&=&\frac{\alpha_s}{2 \pi^2} \frac{1}{\vec{k}_\perp^2} C_F
\int \frac{dx}{x} f_1(x) \left [ \frac{ 1+ \xi^2}{ (1-\xi)_+}
+ \delta(1- \xi) \left ( \ln \frac{x_B^2\zeta^2}{\vec{k}_\perp^2} -1 \right ) \right ] ,
\\
g_{1L}(x_B,k_\perp)&=&\frac{\alpha_s}{2 \pi^2} \frac{1}{\vec{k}_\perp^2} C_F
\int \frac{dx}{x} g_{1L}(x) \left [ \frac{ 1+ \xi^2}{ (1-\xi)_+}
+ \delta(1- \xi) \left ( \ln \frac{x_B^2\zeta^2}{\vec{k}_\perp^2} -1 \right ) \right ] ,
\\
h_1(x_B,k_\perp)&=&\frac{\alpha_s}{2 \pi^2} \frac{1}{\vec{k}_\perp^2} C_F
\int \frac{dx}{x} f_1(x) \left [ \frac{2 \xi}{ (1-\xi)_+}
+ \delta(1- \xi) \left ( \ln \frac{x_B^2\zeta^2}{\vec{k}_\perp^2} -1 \right ) \right ] ,
\end{eqnarray}
where the color factor $C_F=(N_c^2-1)/2N_c$ with $N_c=3$,
$\xi=x_B/x$ and $\zeta^2=(2v \cdot P)^2/v^2$. Here, we have adopted an off-light-cone
vector $v$ to regulate the light-cone singularity associated with the above
calculations~\cite{JiMaYu04}.

In the same spirit, the naive-time-reversal-odd TMD distributions, the quark Sivers function
$f_{1T}^\perp$ and Boer-Mulders function $h_1^\perp$ at large $k_\perp$
can be calculated perturbatively.
The contributions come from the twist-three correlation matrix elements
$T_F$,  $\tilde{T}_F$, and $T_F^{(\sigma)}$.
Furthermore, it is known that the time-reversal odd TMD
distributions is process dependent because the difference on the gauge link directions
will lead to a sign difference between the
SIDIS and Drell-Yan processes\cite{Brodsky:2002cx,Collins:2002kn,Ji:2002aa},
\begin{equation}
f_{1T}^\perp|_{\rm DY}=-f_{1T}^\perp|_{\rm DIS}  , \ h_{1}^\perp|_{\rm DY}=-h_{1}^\perp|_{\rm DIS} \ .
\end{equation}
The quark Sivers function and Boer-Mulders function have been
calculated~\cite{jqvy,Zhou:2008fb}\footnote{The derivative terms in these
results~\cite{jqvy} have been transformed into the non-derivative terms by
partial integrals. The associated boundary
terms were canceled out by the same boundary terms from the derivative terms~\cite{Vogelsang:2009pj}.},
\begin{eqnarray}
f_{1T}^\perp|_{\rm DY}(x_B,k_\perp)&=&\frac{\alpha_s}{\pi} \frac{M^2}{(\vec{k}_\perp^2)^2}
\int \frac{dx}{x} \left [ A_{f_{1T}^\perp}
+ C_F T_F(x,x)\delta(1- \xi) \left ( \ln \frac{x_B^2\zeta^2}{\vec{k}_\perp^2} -1 \right ) \right ]  ,
\\
h_{1}^\perp|_{\rm DY}(x_B,k_\perp)&=&\frac{\alpha_s}{\pi} \frac{M^2}{(\vec{k}_\perp^2)^2}
\int \frac{dx}{x} \left [ A_{h_{1}^\perp}
+ C_F T_F^{(\sigma)}(x,x)\delta(1- \xi) \left ( \ln \frac{x_B^2\zeta^2}{\vec{k}_\perp^2} -1 \right ) \right ]  ,
\end{eqnarray}
where the $A$ factors are defined as
\begin{eqnarray}
A_{f_{1T}^\perp}&=&C_F
 T_F(x,x)\frac{1+\xi^2}{(1-\xi)_+} +
 \frac{C_A}{2}\left[\frac{1+\xi}{1-\xi} T_F(x,x_B)- \frac{1+\xi^2}{1-\xi}T_F(x,x)\right]+ \frac{C_A}{2} \tilde{T}_F(x_B,x) \  ,
\\
A_{h_{1}^\perp}&=&C_F
T_F^{(\sigma)}(x,x)\frac{2 \xi}{(1-\xi)_+}
+\frac{C_A}{2} \left[\frac{2}{1-\xi}T_F^{(\sigma)}(x,x_B)-\frac{2\xi}{1-\xi}T_F^{(\sigma)}(x,x)\right] \  ,
\end{eqnarray}
where the color-factor $C_A=N_c$.

From the above results, we can see that the large transverse momentum
TMD quark distributions have a generic structure. They contain
two parts: one part is similar to the splitting kernel for the relevant
collinear functions, and one term is a delta function associated with a
large logarithm $\ln \zeta^2/k_\perp^2$ which comes from
the light-cone singularity regulated by an off-light cone gauge link
discussed above.
The splitting kernel may be different for different TMD quark distributions.
However, the logarithmic term is the same for all of them. This is because
this term is related to the soft gluon radiation and is spin-independent.
We can also use this as an important consistent check
for all the calculations. In the next section, we will calculate the two $k_\perp$-odd but
time-reversal even quark distributions $g_{1T}$ and $h_{1L}$, and we will
find that they have the same structure.

The energy dependence of the TMD quark distributions, the derivative respected
to $\zeta^2$,  is controlled by the so-called Collins-Soper evolution
equation~\cite{Collins:1981uw,{Idilbi:2004vb}}.
These evolution equations can be used to perform soft-gluon resummation for the
final $k_\perp$ distribution in the cross section and the TMD quark distributions.
It is more convenient to study this resummation in the impact parameter
$b$-space~\cite{Collins:1981uw,{Idilbi:2004vb}}.
We will address this issue in the future.

\section{Transverse momentum dependent quark distributions $g_{1T}$ and $h_{1L}$}

In this section, we will calculate the large transverse momentum behavior for
the two naive-time-reversal-even but $k_\perp$-odd quark distributions:
$g_{1T}$ and $h_{1L}$. These calculations will follow the previous calculations
on the naive-time-reversal-odd quark distributions $f_{1T}^\perp$ and
$h_1^\perp$. However, they differ in a significant way. As shown
in~\cite{Brodsky:2002cx,Collins:2002kn,{Ji:2002aa}},
for the naive-time-reversal-odd distributions, the gauge link contributions play very
important roles. For example, these time-reversal-odd distributions will
have opposite signs between SIDIS and Drell-Yan processes, because
the gauge links go different directions. In the practical calculations, we have
to take pole contributions from the gauge links in these TMD quark
distributions~\cite{jqvy,{Zhou:2008fb}}. The calculations for $g_{1T}$ and $h_{1L}$ are different. Because
they are naive-time-reversal-even, we do not take the pole contributions from the
gauge links. That makes the calculations a little more involved, as
the two-variable dependent correlation functions will enter explicitly.
As we mentioned above, these two TMD  quark distributions will depend on
the correlation functions, $G_D$ ($T_F$), $\tilde G_D$ ($\tilde T_F$),
and $H_D$ ($\tilde T_F^{(\sigma)}$). Moreover, they will have contributions from
the twist-three function $\tilde g$ and $\tilde h$. In the following, we will calculate
these contributions.

\begin{figure}[t]
\begin{center}
\includegraphics[width=9cm]{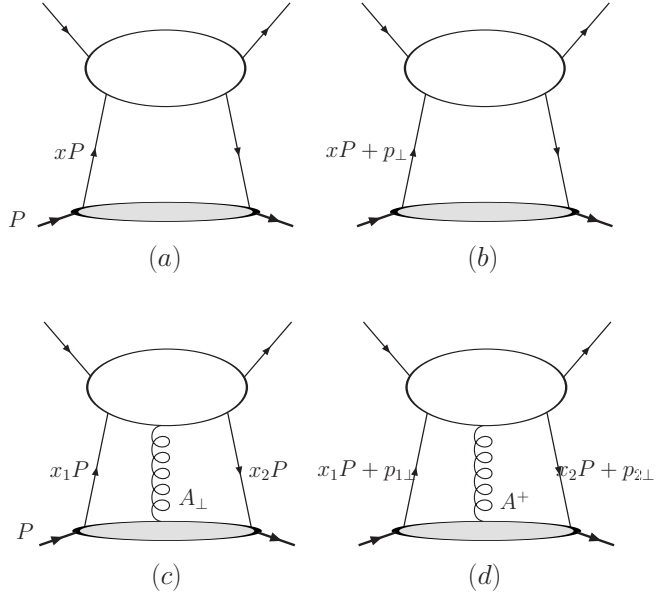}
\end{center}
\vskip -0.4cm \caption{\it Generic diagram interpretations for the twist expansions
in the high energy scattering amplitudes up to
twist-three level: (a) corresponds to a leading twist matrix
element $\langle \bar \psi \psi\rangle$; (b)-(d) for twist-three contributions,
(b) for $\langle \bar \psi \partial_\perp\psi\rangle$, (c) for $\langle \bar \psi A_\perp \psi\rangle$, and
(d) for $\langle \bar \psi \partial_\perp A^+\psi\rangle$. Additional $A^+$ gluon connection
between hard partonic part and the non-perturbative nucleon structure part can be added
to these diagrams.  This is because they do not change the power counting in these diagrams.
The contributions from these diagrams (b-d) are not gauge invariant individually.
However, they will combine into the gauge invariant results in terms of the correlation
functions introduced in Sec.II.} \label{fig1}
\end{figure}

The twist expansion will be the key step in the calculations. The technique
used to calculate these contributions has been well developed in the last few
decades~\cite{Ellis:1982cd,qiu,jqvy,Eguchi:2006qz,new,
Ratcliffe:1985mp,Ji:2001bm,Zhou:2008fb,Yuan:2009dw,Jaffe:1991ra}.
In the following, we will sketch the main points of our calculations
for the TMD quark distributions $g_{1T}$ and $h_{1L}$ and the Drell-Yan lepton
pair production cross section in the next section.

In the twist expansion,
a set of non-perturbative matrix elements of the hadron state will be analyzed
according to the power counting of the associated contributions.
At the twist-three order,
from a generic power counting we have contributions from the following matrix elements
~\cite{Ellis:1982cd,Qiu:1990xxa,Ji:2001bm},
\begin{equation}
\langle \bar \psi\partial_\perp \psi\rangle, ~~\langle \bar \psi A_\perp \psi\rangle,
~~\langle \bar \psi\partial_\perp A^+ \psi\rangle\ ,
\end{equation}
where the quark field spin indices have been suppressed for simplicity.
The relevant Feynman diagrams can be drawn accordingly.
We illustrate the typical diagrams for the associated contributions from the
above matrix elements in Fig.~1. For comparison, we have also shown
the diagram corresponding to the leading-twist contribution from the
matrix element $\langle \bar\psi\psi\rangle$ in Fig.~1a. Figs.~1b-d represent
the contributions up to twist-three quark-gluon correlation matrix elements.
Fig.~1b corresponds to the contributions from the matrix element
$\langle \bar \psi\partial_\perp \psi\rangle$, Fig.~1c from $\langle \bar \psi A_\perp \psi\rangle$,
and Fig.~1d from $\langle \bar \psi\partial_\perp A^+ \psi\rangle$.
Because of additional gluon component in the matrix elements for Fig.~1c and d,
there will be gluon attachment from the nonperturbative part to the
perturbative part as shown in these diagrams.
To calculate the contributions
from Fig.~1b and d, we have to do collinear expansion of the partonic scattering
amplitudes in terms of $p_\perp^\alpha$ and $k_{g\perp}^\alpha=p_{2\perp}^\alpha-p_{1\perp}^\alpha$, respectively.
These expansions, combining with the quark field and gluon field, will lead to the
contributions in terms of the matrix elements: $\langle \bar \psi\partial_\perp \psi\rangle$,
and $\langle \bar \psi\partial_\perp A^+ \psi\rangle$. The calculation of Fig.~1b is
straightforward, without expansion in terms of the transverse momenta of the quarks
and gluon. Furthermore, all these calculations have to be combined into
the gauge invariant matrix elements, such as $G_D$,
$\tilde G_D$, $H_D$, $E_D$, $T_F$, $\tilde T_F$, $T_F^{(\sigma)}$,
$\tilde T_F^{(\sigma)}$, $\tilde g$, and $\tilde h$.

However, these functions
form an over-complete set of correlation functions at this order (twist-three)
as we discussed in Sec.~II. Therefore, we can express the results
in terms of either $D$-type or $F$-type correlation functions.
For example, in the calculations of the twist-three $g_T$ structure
function~\cite{Ratcliffe:1985mp,{Ji:2001bm}}, one has to combine the
contributions from $\langle \bar \psi\partial_\perp \psi\rangle$
and $\langle \bar \psi A_\perp \psi\rangle$ into the gauge invariant form
$\langle \bar \psi D_\perp \psi\rangle$
 which is associated with $G_D$ and $\tilde G_D$.
Meanwhile, for the single spin asymmetry observables (or
the naive-time-reversal-odd TMD quark distributions)~\cite{qiu,new,jqvy,Zhou:2008fb,{Eguchi:2006qz}},
it is more convenient to calculate
the contributions in terms of $\langle \bar \psi\partial_\perp A^+ \psi\rangle$ matrix element.
Such matrix element is part of the gauge invariant matrix element
 $\langle \bar \psi F^{\perp+} \psi\rangle$ which is
associated with the twist-three correlation functions $T_F$, $\tilde T_F$,
and $T_F^{(\sigma)}$. The contributions from the diagrams associated with
$\langle \bar \psi\partial^+ A_\perp \psi\rangle$ have also been shown to exactly
coincide with those from $\langle \bar \psi\partial_\perp A^+ \psi\rangle$ to form
a complete result into the form in terms of $\langle \bar \psi F^{\perp+} \psi\rangle$~\cite{Eguchi:2006qz}.

In the above two examples, it seems that the $\tilde g$ and $\tilde h$ functions
are not necessary in these calculations, because they do not appear in the final
results. This is only because in these calculations one has chosen a particular set of correlation functions
for the final results. Otherwise, $\tilde g$ and $\tilde h$ functions will
show up if we choose different set of correlation functions. For example,  the
structure function $g_T$ can be solely expressed in terms of $G_D$ and $\tilde G_D$.
However, if we rewrite $g_T$ structure function in terms of $T_F$ and $\tilde T_F$,
we will have to introduce the $\tilde g$ function, because of the relation of Eq.~(13).
Similar arguments apply to the single spin asymmetry calculations.
Furthermore, the necessary of $\tilde g$ and $\tilde h$ will be manifest in the
following calculations of the TMD quark distributions $g_{1T}$ and $h_{1L}$.
As shown below, their roles become so essential that we have to include them in the first place.
This can also be seen from the relations between $g_{1T}$ ($h_{1L}$) and
$\tilde g$ ($\tilde h$) discussed  in Sec.~II.

\begin{figure}[t]
\begin{center}
\includegraphics[width=12cm]{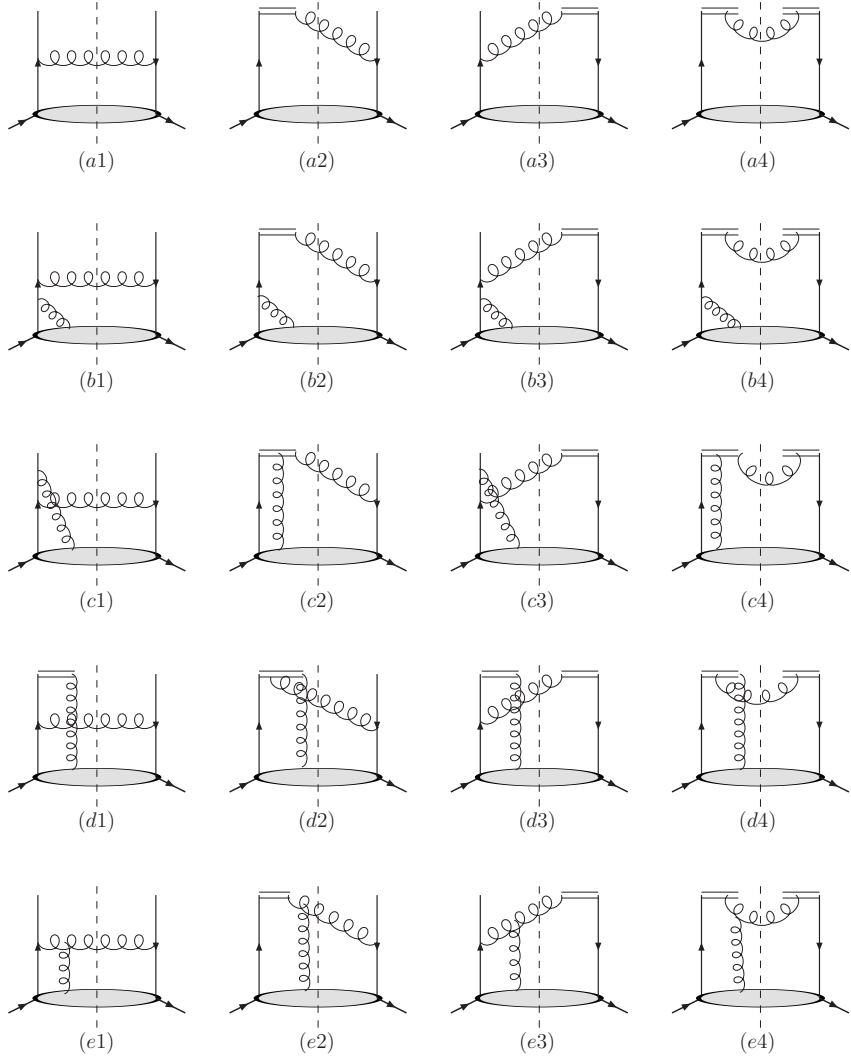}
\end{center}
\vskip -0.4cm \caption{\it Feynman diagrams for the TMD quark distributions at
large transverse momentum calculated from the twist-three quark-gluon
correlation functions. The mirror diagrams of (b1)-(e4) are not shown,
but included in the final results. (a1)-(a4) correspond to the contributions
from the matrix elements of $\langle \bar \psi \partial_\perp\psi\rangle$; (b1)-(b4),
(e1)-(e4), (c1) and (c3) correspond to the diagrams
contributions from $\langle \bar \psi A_\perp \psi\rangle$; and (b1)-(e4)
for $\langle \bar \psi \partial_\perp A^+\psi\rangle$. } \label{fig2}
\end{figure}

We will take $g_{1T}$ calculation as an example to show how we perform the
computation at twist-three level with the quark gluon correlation functions
$G_D$ ($T_F$), $\tilde G_D$ ($\tilde T_F$), and $\tilde g$.
The calculations for the TMD quark distribution $h_{1L}$ and
the Drell-Yan cross sections in the next section will follow accordingly.
As outlined above, we first draw the associated
Feynman diagrams for the large transverse momentum $g_{1T}$ quark
distribution. In order to calculate the large transverse momentum
behavior for the $g_{1T}$ function, we have to radiate a hard gluon. The relevant
diagrams are plotted in Fig.~2, where the probing quark carries the momentum
$k=x_BP+k_\perp$ and the nucleon momentum is denoted by $P$. The double lines
in these diagrams represent the gauge link expansion from the quark distribution
definition in Eq.~(17).
Again, these diagrams include the
contributions from the matrix element $\langle \bar \psi\partial_\perp \psi\rangle$ (a1-a4);
from $\langle \bar\psi A^\mu\psi\rangle$ (b1-e4). To obtain a complete result,
we have to attach the gluon to all possible places as shown in the diagrams
(b1-e4). This also guarantees that we will get the gauge invariant result.
The mirror diagrams of (b1)-(e4) where the gluon attaches to the
right of the cutting line of these diagrams
were not shown in Fig.~2, but included in the final results.
Part of the diagrams (b1)-(e4) correspond
to the contributions from $\langle \bar\psi A_\perp\psi\rangle$, whereas all of them
contribute to that from $\langle \bar\psi \partial_\perp A^+\psi\rangle$.

To calculate the TMD quark distribution $g_{1T}$ at large transverse momentum,
we first compute the individual contributions from the matrix elements shown in
Eq.~(32). Then we will combine the individual results into the gauge invariant twist-three
quark-gluon correlation functions defined in Sec.~II.
We can parameterize the associated matrix elements in Eq.~(32) which correspond
to our calculations. For example, the relevant $\langle \bar\psi \partial_\perp\psi\rangle$
matrix elements are parameterized as
\begin{eqnarray}
M_{ \partial_\perp}^\mu(x) &\equiv& \int \frac{dy^-}{2 \pi}
e^{-ixP^+ y^-} \langle P,S | \bar{\psi}(y^-)
i\partial_\perp^\mu\psi(0) | P,S \rangle
\nonumber\\
&=& \frac{M}{2} \left [
T_{\partial_\perp}(x)i\epsilon^{\mu\nu}_\perp S_{\perp \nu}
p\!\!\!/ +\tilde{T}_{\partial_\perp}(x)S_\perp^\mu \gamma_5
p\!\!\!/ \right ] \ ,
\end{eqnarray}
where $T_{\partial_\perp}$ and $\tilde T_{\partial_\perp}$ correspond to the parts in the
gauge invariant functions $G_D$ and $\tilde G_D$, respectively. Similarly, we can define
the associated $\langle \bar \psi A_\perp\psi\rangle$ matrix elements,
\begin{eqnarray}
M_{ A_\perp}^\mu(x,x_1) &\equiv& \int \frac{dy^-}{2 \pi} \frac{dy_1^-}{2\pi} e^{-ixP^+ y^-} e^{i(x_1-x)P^+y^-_1}
\langle P,S | \bar{\psi}(y^-)  gA_\perp^\mu(y_1^-)\psi(0) | P,S \rangle
\nonumber\\
&=&
\frac{M}{2P^+} \left [ T_{A_\perp}(x,x_1)i\epsilon^{\mu\nu}_\perp S_{\perp \nu} p\!\!\!/
+\tilde{T}_{A_\perp}(x,x_1)S_\perp^\mu \gamma_5 p\!\!\!/ \right ] \ .
\end{eqnarray}
Notice that because of additional gluon attachment to the nucleon state, the above matrix elements
will depend on two variables $(x,x_1)$ which represent the momentum fractions carried
by the quarks from left and right sides of cut line in the diagrams. In the case that there is
no gluon attachment like that in Eq.~(33), they are equal and become one variable.
From the above matrix elements we can easily define those of
$\langle \bar \psi \partial^+A_\perp\psi\rangle$,
\begin{eqnarray}
M_{ \partial^+A_\perp}^\mu(x,x_1) &\equiv& \int
\frac{dy^-}{2 \pi} \frac{dy_1^-}{2\pi} e^{-ixP^+ y^-} e^{i(x_1-x)P^+y^-_1}
\langle P,S | \bar{\psi}(y^-)  g \partial^+A_\perp^\mu (y_1^-)\psi(0) | P,S \rangle
\nonumber\\
&=& \frac{M}{2} \left [ T_{\partial^+A_\perp}(x,x_1) \epsilon^{\nu\mu}_\perp S_{\perp \nu} p\!\!\!/
+\tilde{T}_{\partial^+A_\perp}(x,x_1)iS_\perp^\mu \gamma_5 p\!\!\!/ \right ] \ .
\end{eqnarray}
At the same order, we shall also have the following matrix elements,
\begin{eqnarray}
M_{\partial_\perp A^+}^\mu(x,x_1) &\equiv& - \int \frac{dy^-}{2
\pi} \frac{dy_1^-}{2\pi} e^{-ixP^+ y^-} e^{i(x_1-x)P^+y^-_1}
\langle P,S | \bar{\psi}(y^-)  g \partial_\perp^\mu
A^+(y_1^-)\psi(0) | P,S \rangle
\nonumber\\
&=& \frac{M}{2} \left [ T_{\partial_\perp A^+}(x,x_1) \epsilon^{\nu\mu}_\perp S_{\perp \nu} p\!\!\!/
+\tilde{T}_{\partial_\perp A^+}(x,x_1)iS_\perp^\mu \gamma_5 p\!\!\!/ \right ] \ .
\end{eqnarray}
From the definitions in Sec.~II, we will find that the above matrix elements can form the following
gauge invariant correlation functions,
\begin{eqnarray}
\tilde T_F(x,x_1)&=&\tilde T_{\partial_\perp A^+}(x,x_1)+\tilde T_{\partial^+ A_\perp}(x,x_1) \ ,\\
T_F(x,x_1)&=&T_{\partial_\perp A^+}(x,x_1)+T_{\partial^+ A_\perp}(x,x_1) \ , \\
\tilde g(x)&=&\tilde T_{\partial_\perp}(x)+\int dx_1 P\frac{1}{x-x_1} \tilde T_{\partial_\perp A^+}(x,x_1) \ ,
\end{eqnarray}
where we have used the time-reversal invariance to derive the last equation.
There is no similar relation for $T_{\partial_\perp}$, which on the other hand can be
related to $T_F(x,x)$ depending on the choice of the boundary conditions for
the gauge potential~\cite{Zhou:2009rp},
\begin{equation}
T_{\partial_\perp}(x) =
\left\{
\begin{array}{lll}
{\rm Adv:} & -T_F(x,x), & A_\perp(+\infty)=0 \,\\
{\rm Ret:} & T_F(x,x), & A_\perp(-\infty)=0 \,\\
{\rm PV:}  & 0, & A_\perp(+\infty)+A_\perp(-\infty)=0 \,
\end{array}
\right.
\, .
\end{equation}
It has been shown that the final results on the single spin asymmetries will not
depend on the boundary conditions for the gauge potential, although they correspond
to different relations between the matrix element $T_{\partial_\perp}(x)$ and $T_F(x,x)$,
and different contributions from individual diagrams~\cite{Zhou:2009rp}.

Having sorted out the above relations, it is relative straightforward to perform the calculations.
As outlined above, we will calculate the Feynman diagram contributions in terms of the
matrix elements at the right hand sides of Eqs.~(37-39). These results will be combined
into the gauge invariant correlation functions at the left sides of Eqs.~(37-39). In the following,
we will calculate these contributions separately, and show that how we will combine them
into the gauge invariant results.

\subsection{Contributions from $\tilde T_{\partial_\perp}$ and $T_{\partial_\perp}$}

The contributions from $\tilde T_{\partial_\perp}$ and $T_{\partial_\perp}$ come
from the diagrams Figs.~2(a1-a4). As mentioned above, we will
perform the collinear expansion to calculate their contributions.
That is, the hard partonic part illustrated in the upper parts of these
diagrams can be expanded in terms of the transverse momentum
of the quark connecting to the nucleon state in these diagrams (see also Fig.~1b).
This momentum can be parameterized as
\begin{equation}
p^\mu=xP^\mu+p_\perp^\mu\ ,
\end{equation}
where $x$ is the longitudinal momentum fraction of the nucleon and $p_\perp$
is the transverse momentum. In the collinear expansion, we take the approximation
that $p_\perp\ll k_\perp$, and only keep the leading non-trivial terms which are relevant
for our calculations. For example, the hard partonic part $H$ can be expanded as
\begin{equation}
H(k,p)=H(k,p)|_{p=xP}+p_\perp^\alpha \frac{\partial H(k,p)}{\partial p_\perp^\alpha}|_{p=xP}+\cdots \ ,
\end{equation}
where ellipsis stands for higher order expansion terms,
$\alpha$ is a transverse index $\alpha=1,2$. The first term in the above expansion
does not contribute to the TMD quark distribution $g_{1T}(x_B,k_\perp)$ at large
transverse momentum. The second term will lead to the contribution from the
matrix element $\tilde T_{\partial_\perp}$ and $T_{\partial_\perp}$.

In the calculations, we substitute Eq.~(42) into
the hard partonic part $H(k,p)$ in the Feynman diagrams of Figs.~2(a1-a4),
and take the linear term in the expansion. One particular contribution is
the so-called derivative term, which comes from the expansion of the on-shell
condition for the radiated gluon $k_1=p-k$. To calculate this contribution,
we only keep the $p_\perp$ dependence in the delta function of the on-shell
condition, and set $p_\perp=0$ for all other factors in
the hard partonic scattering amplitude. The final result will be proportional
to the corresponding  Born diagram in the collinear limit~\cite{qiu},
\begin{equation}
g_{1T}(x_B,k_\perp)|_{\tilde T_{\partial_\perp}}^D=
\frac{\alpha_s}{\pi^2}\frac{1}{k_\perp^4}C_F\int\frac{dx}{x}\left(-x\frac{\partial}{\partial x}
\tilde T_{\partial_\perp}(x)\right)(1+\xi^2) \ ,
\end{equation}
where $\xi=x_B/x$ and $1/k_\perp^4$ behavior comes from the power counting
for the $k_\perp$-odd TMD quark distributions.

For the non-derivative terms, we keep all $p_\perp$ dependence in the hard
partonic scattering amplitude, and expand to the linear term in $p_\perp$. Although
it is tedious, the calculation is straightforward, and we obtain
\begin{equation}
\frac{\alpha_s}{\pi^2}\frac{1}{k_\perp^4}C_F\int\frac{dx}{x}
\tilde T_{\partial_\perp}(x)\left[\frac{\xi(1-\xi^2+2\xi)}{(1-\xi)_+}+\delta(1-\xi)
\left(\ln\frac{x_B^2\zeta^2}{k_\perp^2}-1\right)\right] \ ,
\end{equation}
where $\zeta^2$ has been introduced in Sec.~III. After partial integrating for the
derivative term, we can add these two terms Eqs.~(43) and (44)
together\footnote{Note that the boundary
term when we partial integrate the derivative contribution
was canceled out by the boundary term when
we compute the derivative term in Eq.~(43)~\cite{Vogelsang:2009pj}.},
\begin{equation}
g_{1T}(x_B,k_\perp)|_{\tilde T_{\partial_\perp}}=
\frac{\alpha_s}{\pi^2}\frac{1}{k_\perp^4}C_F\int_{x_B}\frac{dx}{x}
\tilde T_{\partial_\perp}(x)\left[\frac{\xi(1+\xi^2)}{(1-\xi)_+}+\delta(1-\xi)
\left(\ln\frac{x_B^2\zeta^2}{k_\perp^2}-1\right)\right] \ .
\end{equation}
Similar calculations can be performed for the contributions from $T_{\partial_\perp}$,
and we find that it does not contribute to the TMD quark distribution $g_{1T}(x_B,k_\perp)$.

\subsection{Contributions from $\tilde T_{A_\perp}$ ($\tilde T_{\partial^+A_\perp}$) and $T_{A_\perp}$ ($T_{\partial^+A_\perp}$)}

Because the attaching gluon is transversely polarized ($A_\perp$), the contributions
from the matrix elements $\tilde T_{A_\perp}$ and $T_{A_\perp}$ come from the
diagrams Figs.~2(b1-b4), (e1-e4), (c1) and (c3). To calculate the contributions
from these diagrams, we take the kinematics illustrated in Fig.~1(c), where
the quark and gluon lines connecting the hard partonic part and the nucleon
state only contain collinear momenta,
\begin{equation}
p_1^\mu=x_1P^\mu,~~p_2^\mu=x_2P^\mu,~~k_g^\mu=(x_2-x_1)P^\mu\ ,
\end{equation}
where $k_g$ is the attaching gluon momentum. These calculations are straightforward,
and we obtain the contributions from $\tilde T_{A_\perp}$,
\begin{eqnarray}
g_{1T}(x_B,k_\perp)|_{\tilde T_{A_\perp}}&=&
\frac{\alpha_s}{\pi^2}\frac{1}{k_\perp^4}\int\frac{dxdx_1}{x}
\tilde T_{A_\perp}(x,x_1)\left\{C_F \left ( \frac{x_B^2}{x^2}+\frac{x_B}{x_1}-\frac{2x_B^2}{x_1x}-\frac{x_B}{x}-1 \right )
\nonumber\right.\\
&&~~\left.+ \frac{C_A}{2}
\frac{(x_B^2+xx_1)(2x_B-x-x_1)}{(x_B-x_1)(x-x_1)x_1}\right\} \ ,
\end{eqnarray}
where we have used the symmetric property for $\tilde T_{A_\perp}(x,x_1)$ to combine
the results from Fig.~2 and their mirrors. Because of additional gluon attachment
in these diagrams, we will have two contributions from two different color factors,
$C_F$ and $C_A$.
Similarly, we have the contribution from $T_{A_\perp}$,
\begin{eqnarray}
g_{1T}(x_B,k_\perp)|_{T_{A_\perp}}&=&
\frac{\alpha_s}{\pi^2}\frac{1}{k_\perp^4}\int{dxdx_1}
 T_{A_\perp}(x,x_1)\left\{ C_F \left(\frac{x_B^2}{x^2}+\frac{x_B}{x_1}-\frac{x_B}{x}-1 \right )
\nonumber\right.\\
&&~~\left.+\frac{C_A}{2}
\frac{x_B^2-xx_1}{(x_1-x_B)x_1} \right\} \ .
\end{eqnarray}
Moreover, the above results can be translated into the
contributions from $\tilde T_{\partial^+A_\perp}$ and $T_{\partial^+A_\perp}$. This is
because we have the following relations between the above matrix elements,
\begin{equation}
\tilde T_{A_\perp}(x,x_1)=P\frac{1}{x-x_1}\tilde T_{\partial^+A_\perp}(x,x_1),~~~
T_{A_\perp}(x,x_1)=P\frac{1}{x-x_1} T_{\partial^+A_\perp}(x,x_1) \  ,
\end{equation}
where the imaginary parts in the right hand sides of the above equations
have been dropped, because they do not contribute to the $g_{1T}$ and $h_{1L}$
calculations here. However, when we calculate the single spin asymmetry
observables (such as the time-reversal-odd Sivers and Boer-Mulders functions),
we have to keep these imaginary parts in the above equations~\cite{Zhou:2009rp}.
Substituting the above results into Eqs.~(47,48), we shall obtain the contributions
from the matrix elements $\tilde T_{\partial^+A_\perp}$ and $T_{\partial^+A_\perp}$.

\subsection{Contributions from $\tilde T_{\partial_\perp A^+}$ and $T_{\partial_\perp A^+}$}

For these contributions, it is the $A^+$ component connecting from the nucleon state
to the hard partonic part, and the gluon can attach to the gauge links in the Feynman diagrams.
Therefore, we will have all diagrams in Figs.~2(b1-e4) contributing to the final results.
Moreover, since these matrix elements involve $\partial_\perp A^+$, we have to perform
the collinear expansion of the hard partonic parts in terms of the gluon transverse momentum.
In doing so, we keep both transverse momenta for the two quark lines connecting the hard
part and the nucleon state,
\begin{equation}
p_1^\mu=x_1P^\mu+p_{1\perp}^\mu,~~~
p_2^\mu=x_2P^\mu+p_{2\perp}^\mu \ .
\end{equation}
Clearly, the kinematics tell us that $k_g^\mu=(x_2-x_1)P^\mu+k_{g\perp}^\mu$
and $k_{g\perp}^\mu=p_{2\perp}^\mu-p_{1\perp}^\mu$. The corresponding collinear
expansion of the hard partonic part takes the following form,
\begin{eqnarray}
H(k;p_1,p_2)&=&H(k,p_1,p_2)|_{p_{1\perp}=p_{2\perp}=0}\nonumber\\
&&+p_{1\perp}^\alpha\frac{\partial H(k;p_1,p_2)}{\partial p_{1\perp}^\alpha}|_{p_{1\perp}=p_{2\perp}=0}
+p_{2\perp}^\alpha\frac{\partial H(k;p_1,p_2)}{\partial p_{2\perp}^\alpha}|_{p_{1\perp}=p_{2\perp}=0}+\cdots \ .
\end{eqnarray}
Again, the expansion coefficients in the above equation can be
calculated following the same method as we discussed in the above
for that for the $\tilde T_{\partial_\perp}$ contribution. For example,
there is also derivative terms associated with $\tilde T_{\partial_\perp A^+}$
matrix element. This contribution also comes from the expansion of
the delta function for the on-shell condition for the radiated gluon $k_1$.
The derivation for this part is similar, and we obtain the following
result,
\begin{eqnarray}
g_{1T}(x_B,k_\perp)|_{\tilde T_{\partial_\perp A^+}}^D
&=&\frac{\alpha_s}{\pi^2}\frac{1}{k_\perp^4}C_F\int\frac{dx}{x}(1+\xi^2)\left(-x\frac{\partial}{\partial x}\int dx_1
P\frac{1}{x-x_1}
\tilde T_{\partial_\perp A^+}(x,x_1)\right) \ , \nonumber\\
\end{eqnarray}
where the same hard coefficient as that for $\tilde T_{\partial}$ appears as it should be
due to the gauge invariance.
As we mentioned above, this derivative term comes from
the delta function expansion for the on-shell condition of $k_1$. To
calculate this contribution, we only keep the $p_{i\perp}$ dependence
in this delta function, and set $p_{i\perp}=0$ for all other factors
in the hard partonic amplitude. Because of this and the fact
that it is the $A^+$ insertion in the hard part, we
can use the Ward identity argument to summarize all diagrams into a simple
factorization form: a product of hard partonic part in the Born diagram
without the gluon insertion and the factor
$1/(x-x_1)$ representing the gluon insertion.

Because they have the same hard coefficient, we can combine the derivative contributions
from $\tilde T_{\partial}$ and $\tilde T_{\partial_\perp A^+}$ together. In particular, by using Eq.~(39), we can add
the results from Eqs.~(43) and (52),
\begin{eqnarray}
g_{1T}(x_B,k_\perp)|_{\tilde g}^D
&=&\frac{\alpha_s}{\pi^2}\frac{1}{k_\perp^4}C_F\int\frac{dx}{x}(1+\xi^2)\left(-x\frac{\partial}{\partial x}\tilde g(x)\right) \ ,
\end{eqnarray}
in term of the gauge invariant correlation function $\tilde g$. This shows how we obtain the gauge
invariant results from the individual contributions. Moreover, it also
demonstrates that $\tilde g$ is an independent
contribution to the large transverse momentum $g_{1T}$ quark distribution.

The non-derivative contributions from $\tilde T_{\partial_\perp A^+}$ can be
calculated accordingly, by keeping all the $p_{i\perp}$ dependence  in
the partonic amplitude. The final result is,
\begin{eqnarray}
g_{1T}(x_B,k_\perp)|_{\tilde T_{\partial_\perp A^+}}&=&
\frac{\alpha_s}{\pi^2}\frac{1}{k_\perp^4}\int\frac{dxdx_1}{x}\frac{1}{x-x_1}\tilde T_{\partial_\perp A^+}(x,x_1)
\left\{C_F\left[\frac{(1-\xi^2+2\xi)\xi}{(1-\xi)_+}\nonumber\right.\right.\\
&&\left.~~+\delta (\xi-1)
\left ( {\rm ln} \frac{x_B^2\zeta^2}{k_\perp^2} -1 \right )\right]+C_F \left ( \frac{x_B^2}{x^2}+\frac{x_B}{x_1}-\frac{2x_B^2}{x_1x}-\frac{x_B}{x}-1 \right )\nonumber\\
&&\left.~~
+ \frac{C_A}{2}
\frac{(x_B^2+xx_1)(2x_B-x-x_1)}{(x_B-x_1)(x-x_1)x_1}
\right\} \ .
\end{eqnarray}
In the above results, there are two terms with the color-factor $C_F$. Clearly, one term
will be combined with that of $\tilde T_{\partial}$ in Eq.~(44) to form the contribution
from the gauge invariant function $\tilde g(x)$, similar to the case for the above derivative contributions.
The other term will be combined with that of $\tilde T_{\partial^+ A_\perp}$ from Eqs.~(47,49)
to form the contribution from the gauge invariant function $\tilde T_{F}(x,x_1)$.

Similarly, we can calculate the contributions from the matrix element $T_{\partial_\perp A^+}$,
\begin{eqnarray}
g_{1T}(x_B,k_\perp)|_{T_{\partial_\perp A^+}}&=&
\frac{\alpha_s}{\pi^2}\frac{1}{k_\perp^4}\int\frac{dxdx_1}{x}\frac{1}{x-x_1} T_{\partial_\perp A^+}(x,x_1)
\left\{C_F \left (\frac{x_B^2}{x^2}+\frac{x_B}{x_1}-\frac{x_B}{x}-1 \right )
\nonumber\right.\\
&&\left.~~
+\frac{C_A}{2}
\frac{x_B^2-xx_1}{(x_1-x_B)x_1}  \right\}\ .
\end{eqnarray}
Again, this result will combine with that from $T_{\partial^+A_\perp}$ from Eq.~(48,49)
to form the gauge invariant result in terms of the gauge invariant function $T_{F}(x,x_1)$.

Combining all these results together, we will obtain the final results for the TMD
quark distribution $g_{1T}(x_B,k_\perp)$ at large transverse momentum,
\begin{eqnarray}
g_{1T}(x_B,k_\perp)=\frac{\alpha_s}{\pi ^2} \frac{M^2}{ (k_\perp^2)^2} \int \frac{dx}{x}
\left \{ A_{g_{1T}} + C_F \tilde{g}(x) \delta (\xi-1) \left ( {\rm ln} \frac{x_B^2\zeta^2}{k_\perp^2} -1 \right ) \right \}
\end{eqnarray}
where $ A_{g_{1T}}$ is given by,
\begin{eqnarray}
A_{g_{1T}}&=&\int dx_1 \left \{ \tilde{g}(x)C_F \left [  \frac{\xi(1+\xi^2)}{(1-\xi)_+} \right ]\delta(x-x_1) \right.\
+P\frac{1}{x-x_1}\tilde{T}_F(x,x_1)
\nonumber\\
&&\times
\left [  C_F \left ( \frac{x_B^2}{x^2}+\frac{x_B}{x_1}-\frac{2x_B^2}{x_1x}-\frac{x_B}{x}-1 \right )
+ \frac{C_A}{2}
\frac{(x_B^2+xx_1)(2x_B-x-x_1)}{(x_B-x_1)(x-x_1)x_1} \right ]
\nonumber\\
&&+ \left .\ \!\!\!
P\frac{1}{x-x_1}T_F(x,x_1)\left [ C_F \left (\frac{x_B^2}{x^2}+\frac{x_B}{x_1}-\frac{x_B}{x}-1 \right )
+\frac{C_A}{2}
\frac{x_B^2-xx_1}{(x_1-x_B)x_1} \right ] \right \} \ ,
\end{eqnarray}
and we have partial integrated the derivative terms.
Using the identities Eqs.~(12) and (13), $A_{g_{1T}}$ can be transformed into the $D$-type correlation
functions,
\begin{eqnarray}
A_{g_{1T}}&=&\int dx_1\left \{ \tilde{g}(x)  \left [ C_F
\frac{1+\xi^2}{(1-\xi)_+}
-\frac{C_A}{2}\frac{1+\xi^2}{1-\xi}\right ]\delta(x-x_1) \right.\
\nonumber\\
&+&
\tilde{G}_D(x,x_1)\left [  C_F \left ( \frac{x_B^2}{x^2}+\frac{x_B}{x_1}-\frac{2x_B^2}{x_1x}-\frac{x_B}{x}-1 \right )
+ \frac{C_A}{2}
\frac{(x_B^2+xx_1)(2x_B-x-x_1)}{(x_B-x_1)(x-x_1)x_1} \right ]
\nonumber\\
&+& \left .\ \!\!\!
G_D(x,x_1) \left [ C_F \left (\frac{x_B^2}{x^2}+\frac{x_B}{x_1}-\frac{x_B}{x}-1 \right )
+\frac{C_A}{2}
\frac{x_B^2-xx_1}{(x_1-x_B)x_1} \right ]\right \} \ .
\end{eqnarray}
This result can also be obtained by directly combining the contributions from $\tilde T_{\partial_\perp}$
in Eq.~(45), $\tilde T_{A_\perp}$ in Eq.~(47), and $T_{A_\perp}$ in Eq.~(48).

For TMD distribution $h_{1L}(x,k_\perp)$, the perturbative calculation follow the similar procedure.
It  receives contributions from $\tilde{T}_F^{(\sigma)}(x,x_1)$,
$\tilde{h}(x)$. We skip the detailed derivation, and list the final result,
\begin{eqnarray}
h_{1L}(x_B,k_\perp)=\frac{\alpha_s}{\pi ^2} \frac{M^2}{ (k_\perp^2)^2} \int \frac{dx}{x}
\left \{ A_{h_{1L}} + C_F \tilde{h}(x) \delta (\xi-1) \left ( {\rm ln} \frac{x_B^2\zeta^2}{k_\perp^2} -1 \right ) \right \}
\end{eqnarray}
where $A_{h_{1L}}$ is defined by,
\begin{eqnarray}
A_{h_{1L}}&=&\int dx_1 \left \{ C_F \left [  \tilde{h}(x) \frac{2
\xi^2}{(1-\xi)_+} \right ]\delta(x_1-x) + P \frac{1}{x-x_1}
\tilde{T}_F^{(\sigma)}(x,x_1) \right .\
\nonumber\\
&&  \times \left .\  \!\!\! \tilde{T}_F^{(\sigma)}(x,x_1)\left [
C_F \frac{2(x-x_1-x_B)}{x_1} + \frac{C_A}{2} \frac{2x_B(x_B x
+x_Bx_1 -x^2-x_1^2)}{(x_B-x_1)(x-x_1)x_1} \right ] \right \} \ .
\end{eqnarray}
Similarly, it can be expressed in terms of the D-type function as,
\begin{eqnarray}
A_{h_{1L}}&=&\int dx_1 \left \{
\tilde{h}(x) \left[C_F\frac{2\xi}{(1-\xi)_+} -\frac{C_A}{2}\frac{2\xi}{1-\xi}\right] \delta(x_1-x) \right .\
\nonumber\\
&+&  \left .\  \!\!\! H_D(x,x_1)\left [ C_F \frac{2(x-x_1-x_B)}{x_1} +
\frac{C_A}{2}
\frac{2x_B(x_B x +x_Bx_1 -x^2-x_1^2)}{(x_B-x_1)(x-x_1)x_1} \right ]
 \right \}
\ .
\end{eqnarray}
Indeed, the above results for the naive-time-reversal-even but $k_\perp$-odd
distributions also have the same structure as those discussed in Sec.III.

\section{single spin $A_{UT}$ and double spin $A_{LT}$ asymmetries in the Drell-Yan lepton pair production process}

In this section, we will calculate the angular distribution of the lepton pair in the polarized Drell-Yan process,
especially the $A_{UT}$ and $A_{LT}$ asymmetries:
One of the incident hadrons is transverse polarized and another
is unpolarized or longitudinal polarized .
We focus on the lepton pair production in hadronic scattering,
\begin{equation}
H_a+H_b\to \gamma^*+X\to \ell^+\ell^-+X \ ,
\end{equation}
which comes from the virtual photon decays. In the leading
order, virtual photon is produced through quark-antiquark
annihilation process, $q\bar q\to \gamma^*$ in the parton picture~\cite{Drell:1970wh}.
In the rest frame of the lepton pair, we can define two angles~\cite{Collins:1977iv}: one
is the polar angle $\theta$ between one lepton momentum and the incident hadron;
the azimuthal angle $\phi$ is defined as the angle between the hadronic
plane and the lepton plane.
The general formalism has been worked out for the angular distributions in the polarized
Drell-Yan process~\cite{Arnold:2008kf}.
For our calculations of $A_{UT}$ and $A_{LT}$,
one can write down the following general structure of the angular distribution of lepton
pair in the Collins-Soper (CS) frame~\cite{Arnold:2008kf},
\begin{eqnarray}
\frac{d \sigma_{(UT,LT)}}{d^4q d \Omega}&=&\frac{\alpha_{em}^2}{2(2 \pi)^4 S^2 Q^2} \times
\nonumber \\
&&\!\!\!\!\!\!\!\!\! \!\!\!\!\!\!\!\!\!\!\!\!\!\!\!\!\!\!\!\!\!\!\!\!\!\!
\left \{  |S_{aT}| \left [ \sin \phi_a \left (
(1+\cos^2\theta)W_T^{UT}+(1-\cos^2\theta)W_L^{UT}+\sin2\theta\cos\phi W_\Delta^{UT}
+\sin^2\theta \cos2 \phi W_{\Delta\Delta}^{UT} \right ) \right .\ \right .\
\nonumber \\
&& \!\!\!\!\!\! \!\!\!\!\!\!\!\!\!\!\!\! \left .\ +
\cos \phi_a \left (\sin2\theta\sin\phi W_\Delta^{'UT}
+\sin^2\theta \sin 2 \phi W_{\Delta\Delta}^{'UT} \right )  \right ]
\nonumber \\
&&\!\!\!\!\!\!\!\!\! \!\!\!\!\!\!\!\!\!\!\!\!\!\!\!\!\!\!\!\!\!\!
+|S_{aT}| S_{bL} \left [ \cos \phi_a \left (
(1+\cos^2\theta)W_T^{LT}+(1-\cos^2\theta)W_L^{LT}+\sin2\theta\cos\phi W_\Delta^{LT}
+\sin^2\theta \cos2 \phi W_{\Delta\Delta}^{LT} \right ) \right .\
\nonumber \\
&& \!\!\!\!\!\!\!
\left .\ \left .\ +
\sin \phi_a \left (\sin2\theta\sin\phi W_\Delta^{'LT}
+\sin^2\theta \sin2 \phi W_{\Delta\Delta}^{'LT} \right )  \right ] \ \right   \} \ ,
\end{eqnarray}
where the orientation of the transverse polarization of the hadron $a$ is expressed
through the CS-angle $\phi_a$, and $\phi$ and $\theta$ have been introduced
above. In the above expressions, $S_{aT}$ and $S_{bL}$ are the hadron $a$ transverse
polarization vector and hadron $b$ longitudinal polarization vector, respectively.
The angular-integrated cross section
is expressed in terms of the $W_T^{UT}$, $W_L^{UT}$, $W_T^{LT}$ and $W_L^{LT}$ as~\cite{Arnold:2008kf},
\begin{eqnarray}
\frac{d \sigma_{(UT,LT)}}{d^4q }&=&\frac{\alpha_{em}^2}{12 \pi^3 S^2 Q^2} \times
\nonumber \\&&
\left \{  |S_{aT}| \sin \phi_a \left ( 2W_T^{UT}+W_L^{UT} \right )
+|S_{aT}| S_{bL}\cos \phi_a \left ( 2W_T^{LT}+W_L^{LT} \right ) \right \} \ .
\end{eqnarray}
In particular, the structure function $2W_T^{UT}+W_L^{UT}$ has been calculated\cite{jqvy}, which
represents the Sivers contribution to the single transverse spin asymmetry. From the above
expression, we also see that the Sivers contribution is the only contribution to the single
spin asymmetry when we integrate out the lepton angles.

In the following, we will calculate the structure functions $W^{UT}$ and $W^{LT}$
in Eq.~(63) in the intermediate transverse momentum region $\Lambda_{\rm QCD}\ll q_\perp\ll Q$,
and will compare the predictions from the collinear factorization and the transverse momentum
dependent approaches. For the single spin asymmetry $W^{UT}$, we follow the previous
calculations~\cite{jqvy}. The only difference is that the calculation in~\cite{jqvy} is equivalent
to the virtual photon production in single transversely polarized nucleon-nucleon scattering,
whereas we will contract the hadronic tensor to the lepton tensor to obtain the angular distributions
of the lepton pair in this process. However, the technique method is the same. Again, as
discussed in~\cite{jqvy}, we will have soft pole and hard pole contributions~\footnote{The so-called
soft-fermion pole will also contribute to the cross sections~\cite{jqvy,{Koike:2009ge}} which have been
neglected in our calculations. However, these
contributions do not change the conclusions of the consistency between the two approaches~\cite{jqvy}.},
and they will have cancelation in the intermediated transverse momentum region.
For the double spin asymmetry part $W^{LT}$,
we will follow the procedure in Sec. IV to calculate the twist-three contributions
to the angular distributions. The corresponding correlation functions will be
$G_D$ ($T_F$), $\tilde G_D$ ($\tilde T_F$), $\tilde g$, and so on.
The similar Feynman diagrams can be drawn accordingly, which can be organized into the contributions
from the twist-three matrix elements $\langle \psi\partial_\perp\psi\rangle$,
$\langle \psi A_\perp\psi\rangle$, and $\langle \psi\partial_\perp A^+\psi\rangle$.
These contributions are then combined into the contributions for the
gauge invariant correlation functions $G_D$, $\tilde G_D$, $\tilde g$, and so on.
In order to calculate the different terms in the angular distribution Eq.~(62),
we choose to work in the Collins-Soper frame~\cite{Collins:1977iv},
where four orthogonal unit vectors are defined as~\cite{Boer:2006eq,{Berger:2007si}},
\begin{eqnarray}
&&T^\mu=\frac{q^\mu}{\sqrt{q^2}} \ ,
\nonumber \\
&&Z^\mu=\frac{2}{\sqrt{Q^2+Q_\perp^2}} \left [ q_{\bar{p}} \tilde{P}^\mu-q_p \tilde{\bar{P}}^\mu \right ] \ ,
\nonumber \\
&&X^\mu=- \frac{Q}{Q_\perp}  \frac{2}{\sqrt{Q^2+Q_\perp^2}}
\left [ q_{\bar{p}} \tilde{P}^\mu + q_p \tilde{\bar{P}}^\mu \right ] \ ,
\nonumber \\
&&Y_\mu=\epsilon^{\mu \nu \alpha \beta } T_\nu Z_\alpha X_\beta \ ,
\end{eqnarray}
where $q^\mu$ is the virtual photon momentum,$P,\bar{P}$ are the
momentum of two hadrons, and we further define
$\tilde{P}^\mu=[P^\mu-(P \cdot q)/q^2 q^\mu ]/ \sqrt{S}$,
$\tilde{\bar{P}}^\mu=[\bar{P}^\mu-(\bar{P} \cdot q)/q^2 q^\mu ]/ \sqrt{S}$
with $q_p\equiv P \cdot q /\sqrt{S} , q_{\bar{p}}=\bar{P} \cdot / \sqrt{S} $
and $S$ the total hadron center of mass energy square, respectively.
The structure function  can be obtained by contracting the hadronic tensor $W^{\mu \nu}$ with
six symmetric tensors constructed by the four orthogonal vectors~\cite{Boer:2006eq,{Berger:2007si}},
\begin{eqnarray}
&& W_T=\frac{1}{2} (X^\mu X^\nu+Y^\mu Y^\nu) W_{\mu \nu}   \ ,
\nonumber \\&&
W_L= Z^\mu Z^\nu W_{\mu \nu} \ ,
\nonumber \\&&
W_\Delta== -\frac{1}{2} ( Z^\mu X^\nu +Z^\nu X^\mu )W_{\mu \nu} \ ,
\nonumber \\&&
W_{\Delta\Delta}=\frac{1}{2} (-X^\mu X^\nu+Y^\mu Y^\nu)W_{\mu \nu} \
\nonumber \\&&
W_\Delta^{'UT}=-\frac{1}{2} ( Y^\mu Z^\nu +Y^\nu Z^\mu)W_{\mu \nu} \ ,
\nonumber \\&&
W_{\Delta\Delta}^{'UT}=\frac{1}{2} ( Y^\mu X^\nu +Y^\nu X^\mu)W_{\mu \nu} \ ,
\end{eqnarray}
and similar expressions hold for the $UT$ and $LT$ structure functions.

Furthermore, we are interested in the cross section contributions in the intermediate
transverse momentum region,
$ \Lambda_{QCD} \ll Q_\perp \ll Q$. To obtain the leading order contributions,
we only keep the leading terms in $Q_\perp/Q$, and
neglect all higher order terms.
With this power counting expansion, six leading order structure
functions survive and can be simplified as,
\begin{eqnarray}
W_T^{UT}&=&
\frac{\alpha_s M}{\pi Q_\perp^3} \sum_q \frac{e_q^2}{N_c} \int \frac{dx}{x} \frac{dz}{z}
\left \{   A_{f_{1T}^\perp}(x) \delta(1-\hat{\xi})+B_{f_{1T}^\perp}(x)\delta(1-\xi) \right \} \bar{f}(z) \ ,
\\
W_{\Delta\Delta}^{UT}&=&
\frac{\alpha_s M}{\pi Q_\perp^3 } \sum_q \frac{e_q^2}{N_c}\int \frac{dx}{x} \frac{dz}{z}
\left \{  A_{h_1^\perp}(z) \delta(1-\xi)+B_{h_1^\perp}(z) \delta(1-\hat{\xi}) \right \} h_1(x)\ ,
\\
W_{\Delta\Delta}^{'UT}&=&-
\frac{\alpha_s M}{\pi Q_\perp^3 } \sum_q \frac{e_q^2}{N_c}\int \frac{dx}{x} \frac{dz}{z}
\left \{  A_{h_1^\perp}(z) \delta(1-\xi)+B_{h_1^\perp}(z) \delta(1-\hat{\xi}) \right \} h_1(x)\ ,\\
W_T^{LT}&=&-
\frac{\alpha_s M}{\pi^2 Q_\perp^3 } \sum_q \frac{e_q^2}{N_c}\int \frac{dx}{x} \frac{dz}{z}
\left \{  A_{g_{1T}^\perp}(x) \delta(1-\hat{\xi})+B_{g_{1T}^\perp}(x)\delta(1-\xi) \right \} \bar{g_1}(z)\ ,
\\
W_{\Delta\Delta}&=&
\frac{\alpha_s M}{\pi^2 Q_\perp^3 } \sum_q \frac{e_q^2}{N_c} \int \frac{dx}{x} \frac{dz}{z}
\left \{  A_{h_{1L}}(z) \delta(1-\xi)+B_{h_{1L}}(z) \delta(1-\hat{\xi}) \right \} h_1(x) \ .
\\
W_{\Delta\Delta}^{'LT}&=&
\frac{\alpha_s M}{\pi^2 Q_\perp^3 } \sum_q \frac{e_q^2}{N_c} \int \frac{dx}{x} \frac{dz}{z}
\left \{  A_{h_{1L}}(z) \delta(1-\xi)+B_{h_{1L}}(z) \delta(1-\hat{\xi}) \right \} h_1(x) \ ,
\end{eqnarray}
where $\xi=x_B/x$, $\hat\xi=z_B/z$, $x_B$ and $z_B$ are defined as
$x_B=Q/\sqrt{S} e^y$ and $z_B=Q/\sqrt{S} e^{-y}$ with $y$ the rapidity of the lepton pair
in the center of mass frame, respectively.
The functions $ A_{f_{1T}}$, $A_{h_1^\perp}$, $A_{g_{1T}}$, $A_{h_{1L}}$ have
been defined in Sec. III with appropriate variable replacements. The functions
$B_{f_{1T}^\perp}$, $B_{h_1^\perp}$, $B_{g_{1T}}$, $B_{h_{1L}}$
are given by,
\begin{eqnarray}
B_{f_{1T}^\perp}(x)&=&C_F T_F(x,x) \left [ \frac{1+\hat{\xi}^2}{(1-\hat{\xi})_+} +2 \delta(\hat{\xi}-1)
\rm{ln}\frac{Q^2}{Q_\perp^2} \right ] \ ,
\\
B_{h_1^\perp}(z)&=&C_F T_F^{(\sigma)}(z,z) \left [ \frac{2\xi}{(1-\xi)_+} +2 \delta(\xi-1)
\rm{ln}\frac{Q^2}{Q_\perp^2} \right ] \ ,
\\
B_{g_{1T}}(x)&=&C_F \tilde{g}(x) \left [ \frac{1+\hat{\xi}^2}{(1-\hat{\xi})_+} +2 \delta(\hat{\xi}-1)
\rm{ln}\frac{Q^2}{Q_\perp^2} \right ] \ ,
\\
B_{h_{1L}}(z)&=&C_F \tilde{h}(z) \left [ \frac{2\xi}{(1-\xi)_+} +2 \delta(\xi-1)
\rm{ln}\frac{Q^2}{Q_\perp^2} \right ] \ ,
\end{eqnarray}
respectively.

Meanwhile, the transverse momentum dependent factorization can be applied in the small transverse momentum,
$Q_\perp \ll Q$. The relevant structure functions can be written as~\cite{Arnold:2008kf},
\begin{eqnarray}
W_T^{UT}&=&\int \frac{\hat{q}_\perp \cdot k_{a\perp} }{M_a} f^\perp_{1T}(x_B,k_{a\perp}) \bar{f}_1(z_B,k_{b\perp}) \ ,
\\
\frac{W_{\Delta\Delta}^{'UT}-W_{\Delta\Delta}^{UT}}{2}&=&-\int \frac{\hat{q}_\perp \cdot k_{b\perp} }{M_a}
h_1(x_B,k_{a\perp}) \bar{h}_1^\perp(z_B,k_{b\perp}) \ ,
\\
W_T^{LT}&=&-\int \frac{\hat{q}_\perp \cdot k_{a\perp} }{M_a} g_{1T}(x_B,k_{a\perp}) \bar{g}_{1L}(z_B,k_{b\perp}) \ ,
\\
\frac{W_{\Delta\Delta}^{'LT}+W_{\Delta\Delta}^{LT}}{2}&=&\int \frac{\hat{q}_\perp \cdot k_{b\perp} }{M_a}
h_1(x_B,k_{a\perp}) \bar{h}_{1L}(z_B,k_{b\perp}) \ .
\end{eqnarray}
where the simple integral symbol represents a complicated integral
$$ \int=\frac{1}{N_c} \sum_q e_q^2 \int d^2k_{a\perp} d^2k_{b\perp} d^2 \lambda_\perp
\left ( S(\lambda_\perp) \right )^{-1} H(Q^2) \delta^2( k_{a\perp}+k_{b\perp}+\lambda_\perp-q_\perp) \ , $$
and $ S(\lambda_\perp) $, $H(Q^2) $ are the soft factor and hard factor, respectively.
In addition, the combinations of structure functions $\frac{W_{\Delta\Delta}^{UT}+W_{\Delta\Delta}^{'UT}}{2}$,
$\frac{W_{\Delta\Delta}^{LT}-W_{\Delta\Delta}^{'LT}}{2}$ also receive the leading power contribution
from the product of TMD distributions $h^\perp_{1T} \times \bar{h}^\perp_1$ and $h^\perp_{1T} \times \bar{h}^\perp_{1L}$
respectively, which is however beyond the scope of the present paper.

When the transverse momenta $k_{a\perp}$ and $k_{b\perp}$ are of order $\Lambda_{QCD}$,
the TMD distribution functions are
entirely non-perturbative objects. But in the large transverse
momentum region $k_{a,b \perp}\gg \Lambda_{QCD}$,
we can calculate the transverse momentum dependence, as shown in Sec. III,
To compare to the results from the collinear factorization calculation, we let one of the
transverse momenta $k_{a\perp}$ , $k_{b\perp}$ , $\lambda_\perp$ be of order $q_\perp$, and
the others are much smaller. After integrating the delta function, we will obtain
\begin{eqnarray}
W_T^{UT}&=&
\frac{\alpha_s M}{\pi Q_\perp^3} \sum_q \frac{e_q^2}{N_c} \int \frac{dx}{x} \frac{dz}{z}
\left \{   A_{f_{1T}^\perp}(x) \delta(1-\hat{\xi})+B_{f_{1T}^\perp}(x)\delta(1-\xi) \right \} \bar{f}(z) \ ,
\\
\frac{W_{\Delta\Delta}^{'UT}-W_{\Delta\Delta}^{UT}}{2}&=&-
\frac{\alpha_s M}{\pi Q_\perp^3 } \sum_q \frac{e_q^2}{N_c}\int \frac{dx}{x} \frac{dz}{z}
\left \{  A_{h_1^\perp}(z) \delta(1-\xi)+B_{h_1^\perp}(z) \delta(1-\hat{\xi}) \right \} h_1(x)\ ,
\\
W_T^{LT}&=&-
\frac{\alpha_s M}{\pi^2 Q_\perp^3 } \sum_q \frac{e_q^2}{N_c}\int \frac{dx}{x} \frac{dz}{z}
\left \{  A_{g_{1T}^\perp}(x) \delta(1-\hat{\xi})+B_{g_{1T}^\perp}(x)\delta(1-\xi) \right \} \bar{g_1}(z)\ ,
\\
\frac{W_{\Delta\Delta}^{'LT}+W_{\Delta\Delta}^{LT}}{2}&=&
\frac{\alpha_s M}{\pi^2 Q_\perp^3 } \sum_q \frac{e_q^2}{N_c} \int \frac{dx}{x} \frac{dz}{z}
\left \{  A_{h_{1L}}(z) \delta(1-\xi)+B_{h_{1L}}(z) \delta(1-\hat{\xi}) \right \} h_1(x) \ .
\end{eqnarray}
It is evident that the above results reproduce the differential cross sections
we derived in the collinear factorization framework.

\section{summary}
In this paper, we have calculated the naive-time-reversal-even but $k_\perp$-odd
TMD distributions $g_{1T}(x,k_\perp)$,
$h_{1L}(x, k_\perp)$ at large transverse momentum, and they are related to a
class of collinear twist-three matrix elements. We further studied
the angular distribution of the lepton pair produced in the polarized Drell-Yan
process for the single spin asymmetry $A_{UT}$ and double spin asymmetry
$A_{LT}$ using the higher twist collinear approach. By comparing these
results with those from the transverse momentum dependent approach,
we found that they are consistent in the intermediated transverse momentum region.

These calculations are not straightforward extensions of the previous calculations
for the naive-time-reversal-odd TMD quark distributions~\cite{jqvy}. This is because,
in the previous case, a pole contribution has to be taken in the final results, which
will  simplify the calculations. In this paper, we have to deal with more complicated
kinematics, similar to next-to-leading order perturbative calculations for the
$g_T$ structure function~\cite{Ji:2001bm}.
To carry out the calculations, we have set up the twist expansion
framework, and in particular, we have shown that the contributions
from twist-three matrix elements will combine into the gauge invariant
form. This shall encourage further developments in the higher-twist
calculations.
For example, an extension to calculate the remaining TMD distributions
$h_{1T}^\perp(x,k_\perp) $ would be possible, though it will be
more complicated because it is related the twist four
collinear matrix element.
The formalism we developed in this paper
can also be extended to other semi-inclusive processes,
such as the semi-inclusive deep inelastic scattering
and back-to-back two hadron production
in $e^+e^-$ annihilation processes. We will address these
issues in future publications.

This work was supported in part by the U.S. Department of Energy
under contract DE-AC02-05CH11231 and the National Natural Science
Foundation of China under the approval No. 10525523. We are grateful
to RIKEN, Brookhaven National Laboratory and the U.S. Department of
Energy (contract number DE-AC02-98CH10886) for providing the
facilities essential for the completion of this work. J.Z. is
partially supported by China Scholarship Council.

\end {document}